\documentclass[english]{article}
\usepackage{babel}
\usepackage{amsmath,amssymb}
\usepackage{graphicx}
\usepackage{epsfig}
\usepackage{epic,eepic}
\usepackage{pstricks
}

\newcommand{\be}{\begin{equation}}
\newcommand{\ee}{\end{equation}}
\newcommand{\bea}{\begin{eqnarray}}
\newcommand{\eea}{\end{eqnarray}}

\begin{document}
\title{Detecting Broken $\mathbf{\mathsf{PT}}$-Symmetry}
\author{Stefan Weigert\\
        Department of Mathematics, University of York\\
        Heslington, York YO10 5DD United Kingdom\\
        \texttt{slow500@york.ac.uk}
       }
\date{February 2006}
\maketitle
\begin{abstract}
A fundamental problem in the theory of $\mathbf{\mathsf{PT}}$-invariant
quantum systems is to determine whether a given system ``respects''
this symmetry or not. If not, the system usually develops non-real eigenvalues. It is shown in this contribution how to algorithmically detect the existence of complex eigenvalues for a given PT-symmetric matrix. The procedure uses classical results from stability theory which qualitatively locate the zeros of real polynomials in the complex plane. The interest and value of the present approach lies in the fact that it avoids diagonalization of the Hamiltonian at hand. 
\end{abstract}

\section{Motivation\label{sec:Motivation}}
When dealing with a non-hermitean operator such as
\begin{equation}
\mathsf{H}=p^{2}+ix^{3},\label{Bessis Hamiltonian}
\end{equation}
one needs to address two questions which do not arise for a Hermitean
operator:
\begin{description}
\item [Q1:]Is the operator $\mathsf{\mathsf{H}}$ is diagonalizable?
\item [Q2:]Does the operator $\mathbf{\mathsf{H}}$ have real eigenvalues
only?
\end{description}
For a randomly picked non-hermitean operator the answers to both questions are unlikely to be positive: it will neither have a complete set of 
eigenfunctions nor a real spectrum. However, operators with $\mathbf{\mathsf{PT}}$-symmetry \cite{bender+98}, 
\begin{equation}
\left[\mathbf{\mathsf{H}},\mathbf{\mathsf{PT}}\right]=0\,,
\label{eq: PT invariance}
\end{equation}
invariant under simultaneous application of parity $\mathbf{\mathsf{P}}$ and time-reversal $\mathbf{\mathsf{T}}$, behave somewhat `better.' $\mathbf{\mathsf{PT}}$-invariant operators
tend to be diagonalizable but for the rare occurrence of exceptional
points, and each of their eigenvalues must be either real or have
a complex conjugate counterpart. Positive answers to \textbf{Q1} and
\textbf{Q2} are necessary in order to attempt a consistent quantum
mechanical interpretation of the operator $\mathbf{\mathsf{H}}$ since
it can be similar to a hermitean operator only then \cite{scholtz+92}. 

To answer these question for a given $\mathbf{\mathsf{PT}}$-invariant
Hamiltonian is by no means straightforward. It is known, for example,
that the operator $\mathbf{\mathsf{H}}$ in Eq. (\ref{Bessis Hamiltonian})
does have only real eigenvalues \cite{dorey+01} while the (likely)
completeness of its eigenfunctions has, apparently, not yet been established rigorously. Perturbative results allow one to confirm
that the spectrum of an initially hermitean operator such as the Hamiltonian of a particle in an oscillator-type potential remains real if a sufficiently weak \textsf{PT}-symmetric term is added \cite{caliceti+80}. As long as
no degeneracies develop, this approach also makes plausible the existence
of a complete set of eigenfunctions; they are, however, not necessarily
pairwise orthogonal with respect to the standard scalar product in
Hilbert space. Technically, the difficulties are due to the fact that
the cubic term in Eq. (\ref{Bessis Hamiltonian}) is unbounded on
the real axis, and the unperturbed operator does not provide a bound
for it. When restricted to a finite interval, a perturbation such
as $igx^{3}, g \in \mathbb{R},$ \emph{is} bounded, and one can reach general conclusions
when perturbing the hermitean boundary value problem with a non-hermitean
\textsf{PT}-symmetric term. Upon treating the \textsf{PT}-symmetrically
perturbed square-well potential \cite{znojil01} in a Krein space
setting, one can show \cite{langer+04} that its eigenvalues remain real if the perturbation  does not move the (non-degenerate) real eigenvalues far enough along
the real axis to create a degeneracy, which is necessary
for complex eigenvalues to emerge. A similar result also follows by
a non-perturbative approach when a ``slightly'' non-selfadjoint term
is added to a self-adjoint operator as is described in \cite{kato80}. 

More is known for \textsf{PT}-symmetric systems with a \emph{finite}-dimensional state space which are described by complex symmetric matrices $\mathbf{\mathsf{M}}$. Let us consider an example which exhibits the essential features: the \emph{discretized} $\mathbf{\mathsf{PT}}$-symmetric square well \cite{weigert2005}.
%
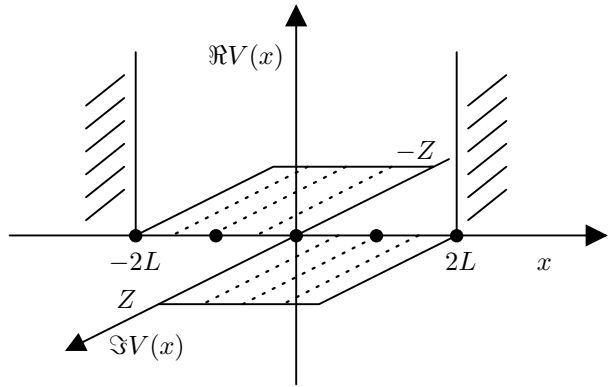
\begin{figure}[ht]
  \begin{center}
\setlength{\unitlength}{0.254mm}
\begin{picture}(312,198)(24,-300)
        \allinethickness{0.254mm}\path(174,-300)(174,-102)\special{sh 1}\path(174,-102)(169,-112)(174,-112)(179,-112)(174,-102) 
        \allinethickness{0.254mm}\path(24,-222)(336,-222)\special{sh 1}\path(336,-222)(326,-217)(326,-222)(326,-227)(336,-222) 
        \allinethickness{0.254mm}\path(254,-182)(54,-282)\special{sh 1}\path(54,-282)(64,-282)(62,-278)(60,-274)(54,-282) 
        \allinethickness{0.254mm}\special{sh 0.99}\put(174,-222){\ellipse{6}{6}} 
        \allinethickness{0.254mm}\special{sh 0.99}\put(132,-222){\ellipse{6}{6}} 
        \allinethickness{0.254mm}\special{sh 0.99}\put(90,-222){\ellipse{6}{6}} 
        \allinethickness{0.254mm}\special{sh 0.99}\put(258,-222){\ellipse{6}{6}} 
        \allinethickness{0.254mm}\special{sh 0.99}\put(216,-222){\ellipse{6}{6}} 
        \allinethickness{0.254mm}\path(90,-126)(90,-222)(162,-186)(246,-186)(244,-186) 
        \allinethickness{0.254mm}\path(102,-258)(186,-258)(258,-222)(258,-126)(258,-126) 
        \allinethickness{0.254mm}\path(264,-154)(284,-138) 
        \allinethickness{0.254mm}\path(264,-166)(284,-150) 
        \allinethickness{0.254mm}\path(264,-178)(284,-162) 
        \allinethickness{0.254mm}\path(264,-202)(284,-186) 
        \allinethickness{0.254mm}\path(264,-214)(284,-198) 
        \allinethickness{0.254mm}\path(264,-190)(284,-174) 
        \allinethickness{0.254mm}\path(64,-154)(84,-138) 
        \allinethickness{0.254mm}\path(64,-166)(84,-150) 
        \allinethickness{0.254mm}\path(64,-178)(84,-162) 
        \allinethickness{0.254mm}\path(64,-190)(84,-174) 
        \allinethickness{0.254mm}\path(64,-202)(84,-186) 
        \allinethickness{0.254mm}\path(64,-214)(84,-198) 
        \put(300,-240){$x$} 
        \put(252,-240){$2L$} 
        \put(76,-240){$-2L$} 
        \put(128,-132){$\Re V(x)$} 
        \put(76,-284){$\Im V(x)$} 
        \put(226,-182){$-Z$} 
        \put(80,-260){$Z$} 
        \allinethickness{0.254mm}\dottedline{6}(111,-221)(180,-186) 
        \allinethickness{0.254mm}\dottedline{6}(131,-221)(200,-186) 
        \allinethickness{0.254mm}\dottedline{6}(155,-221)(224,-186) 
        \allinethickness{0.254mm}\dottedline{6}(146,-258)(215,-223) 
        \allinethickness{0.254mm}\dottedline{6}(127,-257)(196,-222) 
        \allinethickness{0.254mm}\dottedline{6}(169,-257)(238,-222) 
\end{picture}
\caption{Discretized PT-symmetric well: the wave function takes non-zero values at three points $x=0,\pm L$ only (cf. text)}
\label{fig: discretePTwell}
\end{center}
\end{figure}
This model, sketched in Fig. \ref{fig: discretePTwell}, is obtained upon discretizing the configuration space of a particle moving freely between walls at $x=\pm2L$, subjected to a piecewise constant potential $\pm iZ,Z\in\mathbb{R}$. Defining a
wave function which takes values at the points $x=0,\pm L,\pm2L$,
and satisfies ``hard'' boundary conditions at $x=\pm2L$, an effective Hamiltonian is obtained, 
\begin{equation}
\mathbf{\mathsf{H}}=\left(\begin{array}{ccc}
i\xi & 1 & 0\\
1 & 0 & 1\\
0 & 1 & -i\xi\end{array}\right),\quad\xi=2mL^{2}Z/\hbar^{2}\,.\label{eq: PT sym Hamiltonian}
\end{equation}
This matrix is invariant under the action of parity $\mathbf{\mathsf{P}}$, a matrix with ones along the minor diagonal and zeros elsewhere, followed by complex conjugation, overall equivalent to Eq. (\ref{eq: PT invariance}). The eigenvalues of $\mathbf{\mathsf{H}}$ are given by the roots of its characteristic polynomial, 
\begin{equation}
p_{\mathbf{\mathsf{H}}}(\lambda)=\lambda(\lambda^{2}-(2-\xi^{2}))\,,
\label{eq: charpoly Discrete SW}
\end{equation}
reading explicitly,
\begin{equation}
E_{0}=0\;\mbox{{ and }}\; E_{\pm}=\pm\sqrt{2-\xi^{2}}\in\left\{ \begin{array}{rl}
\mathbb{R} & \mbox{ if }|\xi|<\sqrt{2}\, ,\\
i\mathbb{R} & \mbox{ if }|\xi|>\sqrt{2}\, .\end{array}\right.
\label{eq: Discret SW eigenvalues}
\end{equation}
The possibility to analytically determine the eigenvalues of $\mathbf{\mathsf{H}}$
provides immediate and exhaustive answers to both $\textbf{Q1}$ and
$\textbf{Q2}$, summarized briefly now. The zero eigenvalue (with its associated eigenstate) persists for all values of $\xi$, while
the remaining two change their character with varying strength of
the parameter $Z$. Three regions can be identified (cf. Part (b) of Fig.
(\ref{fig: exactvsalgevs})): depending on the magnitude of $\xi$, there 
is either a pair of complex-conjugate or a pair of real eigenvalues. However, the matrix $\mathbf{\mathsf{H}}$ is 
not diagonalizable for $\xi=\pm\sqrt{2}$: only a 
single eigenvector is associated with $E_{\pm}=0$,
while the \emph{algebraic} multiplicity of this eigenvalue is two \cite{weigert2005}. 

For matrices $\mathbf{\mathsf{M}}$ of larger dimensions no analytic
expressions for the eigenvalues exist. To overcome this shortcoming,
an algorithm has been proposed which is capable to detect whether
a $\mathbf{\mathsf{PT}}$-invariant matrix is diagonalizable
or not \cite{weigert06}. The relevant information is coded
in the \emph{minimal} polynomial of the matrix which one can construct
without knowing the eigenvalues of $\mathbf{\mathsf{M}}$, just as
its \emph{characteristic} polynomial. This approach answers \textbf{Q1}
systematically, circumventing the need to numerically calculate the
eigenvalues of $\mathbf{\mathsf{M}}$. This is important from a conceptual
point of view. 

In the present contribution, a second, independent algorithm will be presented which answers \textbf{Q2} for any \textsf{PT}-symmetric matrix. Both the number of its real eigenvalues and
the number of pairs of complex eigenvalues are obtained by manipulating  the coefficients of the characteristic polynomial of $\mathbf{\mathsf{M}}$. This information will be called the \emph{qualitative spectrum} of $\mathbf{\mathsf{M}}$. 

The interest of the method proposed is due to the fact that it is
possible to extract nothing but the desired information about the
eigenvalues, namely their location relative to the real axis in the
complex plane. Problems of this type arise when the stability of dynamical systems is addressed where it is crucial to determine whether the eigenvalues of a given matrix have negative real parts. 

In Section \ref{sec:Stability-and-inertia}, the notion of inertia
is introduced for Hermitean matrices, followed by Jacobi's criterion
of stability for such matrices. Then, the breaking (or not) of $\mathbf{\mathsf{PT}}$-symmetry
is described in terms of a modified inertia. Next, a theorem by Jacobi
and Borhard is presented which locates the zeros of real polynomials
in the complex plane. Section \ref{sec:Alg-detecting-complex-EVs}
combines all this to formulate an algorithm which, given a $\mathbf{\mathsf{PT}}$-invariant
(or quasi-Hermitean) matrix, outputs the number of its real and of its complex eigenvalues. Finally, the algorithm is illustrated by applying it to the discretized $\mathbf{\mathsf{PT}}$-symmetric
square-well potential introduced above, outputting correctly its qualitative spectrum. 

\section{Stability and inertia of matrices \label{sec:Stability-and-inertia}}

Consider a dynamical system which is described exactly or, after some
approximation, by the equation
\begin{equation}
\frac{d\mathbf{x}}{dt}=\mathbf{\mathsf{M}}\cdot\mathbf{x}\,,
\label{eq:dynamical system}
\end{equation}
where $\mathbf{\mathsf{M}}$ is a fixed hermitean (or real symmetric)
matrix of dimensions $(N\times N)$, and the vector $\mathbf{x}(t)$
gives the state of the system at time $t$. In many applications,
one needs to know whether the solutions of Eq. (\ref{eq:dynamical system}) are \emph{stable}: this is the case if all eigenvalues $\mathbf{\mathsf{M}}_{n}$ of $\mathbf{\mathsf{M}}$ have negative real parts,
\begin{equation}
\mathfrak{Re}\,\mathbf{\mathsf{M}}_{n}<0\,.
\label{eq: stability}
\end{equation}
Indeed, no solution of Eq. (\ref{eq:dynamical system})
will grow without bounds if (\ref{eq: stability}) holds, making it possible 
to qualitatively predict the system's long-term behaviour.
Let us characterize a matrix $\mathsf{M}$ by a triple
of non-negative integers, its \emph{inertia} \cite{lancaster+85}
with respect to the imaginary axis, 
\begin{equation}
\texttt{In}\,\mathbf{\mathsf{M}}=\{\nu,\delta,\pi\}\,,
\label{eq: imaginary inertia of M}
\end{equation}
where $\nu$ and $\pi$ are the number of its eigenvalues
with negative and positive real parts, respectively, while $\delta$
counts the eigenvalues on the imaginary axis (cf. (\ref{fig: visualInertia}) for an illustration). A stable matrix $\mathbf{\mathsf{M}}$ has an inertia of the form 
\begin{equation}
\texttt{In}\,\mathbf{\mathsf{M}}=\{ N,0,0\}\,,
\label{eq:stable inertia}
\end{equation}
while a matrix is called \emph{marginally stable} if none of its eigenvalues have a negative real part, allowing for the presence of purely imaginary eigenvalues,
\begin{equation}
\texttt{In}\,\mathbf{\mathsf{M}}=\{ N-m,m,0\}\,,\quad0<m\leq N\,.
\label{eq: marg stable inertia}
\end{equation}
Whenever $\pi>0$, the matrix $\mathbf{\mathsf{M}}$ is called \emph{unstable} since there is at least one solution of (\ref{eq:dynamical system}) which will grow without bound. 

\subsection{Inertia of Hermitean matrices: Jacobi's method\label{sub:Inertia-of-Hermitean}}

Jacobi devised an ingenious method \cite{jacobi57} to determine the inertia of a given
(non-singular) \emph{hermitean} matrix $\mathbf{\mathsf{L}}$ of size $(N\times N)$.
First, calculate the determinants $d_{n}$ of its $N$ leading principal
submatrices $\mathbf{\mathsf{L}}_{1},\mathbf{\mathsf{L}}_{2},\ldots,\mathbf{\mathsf{L}}_{N}\equiv\mathbf{\mathsf{L}}$,
\begin{equation}
d_{n}\equiv\det\mathbf{\mathsf{L}}_{n}\,,\quad n=1,\ldots,N\,,
\label{eq: leading determinants}
\end{equation}
all of which must be different from zero; second, write
down a ``$+$'' followed by the sequence of signs $\sigma_n$ of the $N$ determinants $d_{n}$,
\begin{center}
\begin{equation}
+,\sigma_{1},\sigma_{2},\ldots,\sigma_{N}\,,\qquad\sigma_{n}=\frac{d_{n}}{|d_{n}|}=\pm1\,.
\label{eq:signsequence}
\end{equation}
\end{center}
These $(N+1)$ signs
encode the inertia of the matrix $\mathbf{\mathsf{L}}$: the number
of sign \emph{changes} in this sequence equals the number $\pi$ of
eigenvalues with positive real part, while the number of \emph{constancies} in signs equals the number $\nu$ of its negative eigenvalues:
\begin{equation}
\left.\begin{array}{c}
\#\mbox{ of constancies in}\,(\ref{eq:signsequence})\equiv\pi\\
\#\mbox{ of alterations in}\,(\ref{eq:signsequence})\equiv\nu\end{array}\right\} \,\Rightarrow\,\,\texttt{In}\,\mathbf{\mathsf{L}}=(\nu,0,\pi)\,.
\label{eq: Jacobis rule}
\end{equation}
The matrix $\mathbf{\mathsf{L}}$ cannot have a zero eigenvalue, that is, 
$\delta \equiv 0$, since all leading subdeterminants including $d_{N}$
have been assumed to be nonzero.

The following paragraph will show that it is possible to detect
the location of the eigenvalues of a $\mathbf{\mathsf{PT}}$-symmetric
(hence non-hermitean) matrix relative to the \emph{real} axis by similar
methods.

\subsection{Stability and inertia of $\mathbf{\mathsf{PT}}$-invariant matrices}

\begin{figure}[ht]
\begin{center}
\setlength{\unitlength}{0.00061242in}
\begingroup\makeatletter\ifx\SetFigFont\undefined
\def\x#1#2#3#4#5#6#7\relax{\def\x{#1#2#3#4#5#6}}%
\expandafter\x\fmtname xxxxxx\relax \def\y{splain}%
\ifx\x\y   
\gdef\SetFigFont#1#2#3{%
  \ifnum #1<17\tiny\else \ifnum #1<20\small\else
  \ifnum #1<24\normalsize\else \ifnum #1<29\large\else
  \ifnum #1<34\Large\else \ifnum #1<41\LARGE\else
     \huge\fi\fi\fi\fi\fi\fi
  \csname #3\endcsname}%
\else
\gdef\SetFigFont#1#2#3{\begingroup
  \count@#1\relax \ifnum 25<\count@\count@25\fi
  \def\x{\endgroup\@setsize\SetFigFont{#2pt}}%
  \expandafter\x
    \csname \romannumeral\the\count@ pt\expandafter\endcsname
    \csname @\romannumeral\the\count@ pt\endcsname
  \csname #3\endcsname}%
\fi
\fi\endgroup
{\renewcommand{\dashlinestretch}{30}
\begin{picture}(5623,5638)(0,-10)
\put(2440,5140){\makebox(0,0)[b]{\smash{{{\SetFigFont{10}{12.0}{rm}$\Im M$}}}}}
\put(4750,4150){\blacken\ellipse{120}{120}}
\put(4750,4150){\ellipse{120}{120}}
\put(640,2890){\blacken\ellipse{120}{120}}
\put(640,2890){\ellipse{120}{120}}
\put(5196,4184){\blacken\ellipse{120}{120}}
\put(5196,4184){\ellipse{120}{120}}
\put(5196,1533){\blacken\ellipse{120}{120}}
\put(5196,1533){\ellipse{120}{120}}
\put(4289,1758){\blacken\ellipse{120}{120}}
\put(4289,1758){\ellipse{120}{120}}
\put(4289,4015){\blacken\ellipse{120}{120}}
\put(4289,4015){\ellipse{120}{120}}
\put(3340,3509){\blacken\ellipse{120}{120}}
\put(3340,3509){\ellipse{120}{120}}
\put(2890,3059){\blacken\ellipse{120}{120}}
\put(2890,3059){\ellipse{120}{120}}
\put(2890,2721){\blacken\ellipse{120}{120}}
\put(2890,2721){\ellipse{120}{120}}
\put(3340,2264){\blacken\ellipse{120}{120}}
\put(3340,2264){\ellipse{120}{120}}
\put(3790,1983){\blacken\ellipse{120}{120}}
\put(3790,1983){\ellipse{120}{120}}
\put(3790,3790){\blacken\ellipse{120}{120}}
\put(3790,3790){\ellipse{120}{120}}
\put(2440,2890){\blacken\ellipse{120}{120}}
\put(2440,2890){\ellipse{120}{120}}
\put(1540,2890){\blacken\ellipse{120}{120}}
\put(1540,2890){\ellipse{120}{120}}
\put(1990,2890){\blacken\ellipse{120}{120}}
\put(1990,2890){\ellipse{120}{120}}
\put(1090,2890){\blacken\ellipse{120}{120}}
\put(1090,2890){\ellipse{120}{120}}
\thicklines
\put(5140,5140){\ellipse{500}{500}}
\thinlines
\path(3130.000,358.000)(2890.000,190.000)(3130.000,22.000)
\dottedline{60}(2890,190)(5590,190)
\thicklines
\thinlines
\path(22.000,3130.000)(190.000,2890.000)(358.000,3130.000)
\dottedline{60}(190,2890)(190,5590)
\thicklines
\dottedline{60}(190,190)(2890,190)
\thinlines
\path(2650.000,22.000)(2890.000,190.000)(2650.000,358.000)
\thicklines
\dottedline{60}(190,190)(190,2890)
\thinlines
\path(358.000,2650.000)(190.000,2890.000)(22.000,2650.000)
\path(589,2431)(589,2432)(590,2435)
	(593,2442)(597,2455)(602,2473)
	(609,2494)(618,2517)(626,2541)
	(636,2565)(646,2588)(656,2609)
	(668,2629)(680,2645)(693,2658)
	(710,2665)(722,2661)(729,2649)
	(733,2632)(735,2612)(736,2593)
	(739,2575)(746,2563)(756,2560)
	(770,2568)(780,2581)(789,2599)
	(797,2620)(803,2642)(809,2666)
	(815,2691)(820,2716)(824,2740)
	(828,2762)(831,2781)(835,2806)
\path(863.438,2679.924)(835.000,2806.000)(768.644,2695.091)
\path(2455,614)(2458,614)(2465,615)
	(2476,617)(2492,619)(2510,621)
	(2530,625)(2551,628)(2572,632)
	(2592,637)(2610,642)(2628,648)
	(2643,656)(2654,664)(2661,676)
	(2658,685)(2647,691)(2631,694)
	(2614,696)(2596,698)(2581,702)
	(2571,709)(2570,720)(2580,735)
	(2593,745)(2608,755)(2626,764)
	(2646,772)(2666,780)(2687,787)
	(2709,794)(2731,801)(2752,807)
	(2771,812)(2789,817)(2820,825)
\path(2715.801,748.537)(2820.000,825.000)(2691.813,841.492)
\thicklines
\path(190,2890)(190,865)
\path(2890,190)(865,190)
\path(190,2890)(5590,2890)
\blacken\thinlines
\path(5350.000,2818.000)(5590.000,2890.000)(5350.000,2962.000)(5350.000,2818.000)
\thicklines
\path(2890,190)(2890,5590)
\blacken\thinlines
\path(2962.000,5350.000)(2890.000,5590.000)(2818.000,5350.000)(2962.000,5350.000)
\thicklines
\path(4915,190)(2890,190)
\path(190,4900)(190,2875)
\put(611,2243){\makebox(0,0)[b]{\smash{{{\SetFigFont{10}{12.0}{rm}$\delta_{\Re}$}}}}}
\put(430,1540){\makebox(0,0)[lb]{\smash{{{\SetFigFont{10}{12.0}{rm}$\nu_{\Re}$}}}}}
\put(415,4240){\makebox(0,0)[lb]{\smash{{{\SetFigFont{10}{12.0}{rm}$\pi_{\Re}$}}}}}
\put(5125,5068){\makebox(0,0)[b]{\smash{{{\SetFigFont{10}{12.0}{rm}$M$}}}}}
\put(1540,430){\makebox(0,0)[b]{\smash{{{\SetFigFont{10}{12.0}{rm}$\nu$}}}}}
\put(3930,444){\makebox(0,0)[b]{\smash{{{\SetFigFont{10}{12.0}{rm}$\pi$}}}}}
\put(2320,430){\makebox(0,0)[b]{\smash{{{\SetFigFont{10}{12.0}{rm}$\delta$}}}}}
\put(5111,2483){\makebox(0,0)[b]{\smash{{{\SetFigFont{10}{12.0}{rm}
$\Re M$}}}}}
\thinlines
\put(4750,1567){\blacken\ellipse{120}{120}}
\put(4750,1567){\ellipse{120}{120}}
\end{picture}
}
\caption{Inertia of a $(17\times 17)$ matrix with broken $\mathbf{\mathsf{PT}}$-symmetry; its imaginary and real inertia are given by 
$\texttt{In} \mathbf{\mathsf{M}} = \{ 5,2,10 \}$ and $\texttt{In}_{\Re} \,\mathbf{\mathsf{M}}=\{6,5,6\}$, respectively.}
\label{fig: visualInertia}
\end{center}
\end{figure}
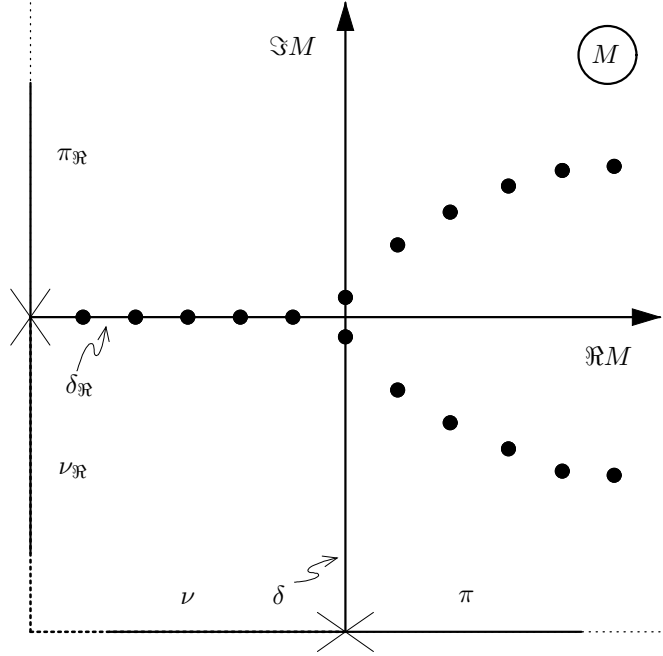

A non-hermitean matrix $\mathbf{\mathsf{H}}$ with $\mathbf{\mathsf{PT}}$-symmetry
satisfies (\ref{eq: PT invariance}) which implies that its characteristic
polynomial
\begin{equation}
p_{\mathsf{H}}(\lambda)=\sum_{n=0}^{N}h_{n}\lambda^{n}
\label{eq: charpoly}
\end{equation}
has real coefficients $h_{n}$ only, 
\begin{equation}
p_{\mathbf{\mathsf{H}}}^{*}(\lambda)=p_{\mathbf{\mathsf{H}}}(\lambda^{*})\,.
\label{eq: real charpoly}
\end{equation}
As a consequence, the zeros of this polynomial are either real or
they come in complex-conjugate pairs. To distinguish between broken
and unbroken $\mathbf{\mathsf{PT}}$-symmetry, it is useful to introduce
the inertia of a matrix $\mathbf{\mathsf{H}}$ with respect to the
\emph{real} axis, 
\begin{equation}
\texttt{In}_{\Re}\,\mathbf{\mathsf{H}}=\{\nu_{\Re},\delta_{\Re},\pi_{\Re}\}\,,
\label{eq: real inertia}
\end{equation}
where the triple $\{\nu_{\Re},\delta_{\Re},\pi_{\Re}\}$ of integers
denotes the number of eigenvalues of $\mathbf{\mathsf{H}}$ with negative,
vanishing, and positive imaginary part (cf. Fig. (\ref{fig: visualInertia})). 
The inertia of a matrix with real eigenvalues only, corresponding to \emph{unbroken}
$\mathbf{\mathsf{PT}}$-symmetry, reads
\begin{equation}
\texttt{In}_{\Re}\,\mathbf{\mathsf{H}}=\{0,N,0\}\,,
\label{eq:unbroken inertia}
\end{equation}
while \emph{broken} $\mathbf{\mathsf{PT}}$-symmetry is
signaled by an inertia of the form 
\begin{equation}
\texttt{In}_{\Re}\,\mathbf{\mathsf{H}}=\{ m,N-2m,m\},\qquad m>0\,,
\label{eq: mixed inertia}
\end{equation}
corresponding to $m$ pairs of complex eigenvalues and $(N-2m)$
real ones. Let us know turn to the question how to determine the real inertia of a matrix with $\mathbf{\mathsf{PT}}$-symmetry. 

\subsection{ Zeros of real polynomials }

Given a real polynomial $p(\lambda)$ of degree $N$, Borhard
\cite{borhardt47} and Jacobi \cite{jacobi57/2} propose to proceed
as follows to obtain the number of its real zeros. To begin,
one determines the first $(2N-2)$ Newton sums associated with the
polynomial $p(\lambda)$ defined by
\begin{equation}
s_{0}=N\,,\qquad s_{n}=\lambda_{1}^{n}+\ldots+\lambda_{N}^{n}\,,\qquad n=1,2,\ldots,2N-2\,.
\label{eq: def of Newton sums}
\end{equation}
This is possible \emph{without} knowing the zeros $\lambda_{1},\ldots,\lambda_{N}$,
since one can \cite{lancaster+85} either define the numbers $s_{n}$ recursively in terms
of the coefficients $h_n$ of the polynomial  or generate
them by means of the identity
\begin{equation}
\frac{dp(\lambda)}{d\lambda}=(s_{0}\lambda^{-1}+s_{1}\lambda^{-2}+\ldots)p(\lambda)\,.
\label{eq: generate newton sums}
\end{equation}
Once the Newton sums have been calculated, one introduces the real symmetric (and Hermitean) matrix
\begin{equation}
\mathfrak{H}_{p}=\left(\begin{array}{ccccc}
s_{0} & s_{1} & s_{2} & \cdots & s_{N-1}\\
s_{1} & s_{2} &  & \cdots & s_{N}\\
s_{2} &  &  &  & s_{N+1}\\
\vdots &  &  &  & \vdots\\
s_{N-1} & s_{N} &  & \cdots & s_{2N-2}\end{array}\right)\,,
\label{eq: hankel of p}
\end{equation}
which is of Hankel type. One can thus apply the method presented in Section \ref{sub:Inertia-of-Hermitean} to determine its imaginary inertia. This is useful since Borhard \cite{borhardt47} and Jacobi \cite{jacobi57/2} have shown\footnote{The content of Refs. \cite{borhardt47} and \cite{jacobi57/2} is
described in \cite{lancaster+85}.} that the inertia of $\mathfrak{H}_{p}$, in fact, encodes the structure
of the zeros of the polynomial $p(\lambda)$:
\begin{equation}
\texttt{In}\;\mathfrak{H}_{p}=\{\nu,0,\pi\}
 \; \Rightarrow \; 
 p(\lambda) \mbox{ has  }
\left\{
\begin{array}{l}
 \pi - \nu \mbox{ different real zeros,}\\
 \nu \mbox{ different pairs of complex-conjugate zeros.} 
\end{array}
\right. 
\label{eq: inertia and poly}
\end{equation}
Let us imagine that the real polynomial $p(\lambda)$ is the characteristic polynomial $p_{\mathbf{\mathsf{H}}}(\lambda)$ associated with a $\mathbf{\mathsf{PT}}$-invariant matrix $\mathbf{\mathsf{H}}$. Then, the result (\ref{eq: inertia and poly}) says that $\mathbf{\mathsf{H}}$ has $\nu$ pairs of different complex eigenvalues and $(\pi - \nu)$ different real eigenvalues if the Hankel matrix $\mathfrak{H}_{\mathbf{\mathsf{H}}}$ associated with $p_\mathbf{\mathsf{H}}(\lambda)$ has $\nu (\pi)$ eigenvalues with negative (positive) real part. Expressed in terms of inertias, this result reads 
\begin{equation}
\texttt{In}\;\mathfrak{H}_{\mathbf{\mathsf{H}}}=\{\nu,0,\pi\}
 \quad \Longrightarrow \quad 
\texttt{In}_\Re\;\mathbf{\mathsf{H}}=\{\nu,\pi-\nu,\nu\}\,.
\label{eq: inertia relations}
\end{equation}
The next Section will collect
the results obtained so far and present them as an algorithm to determine
the number of complex pairs and real eigenvalues of a $\mathbf{\mathsf{PT}}$-invariant matrix. 

\section{ Algorithm detecting complex eigenvalues\label{sec:Alg-detecting-complex-EVs}}

Given a matrix $\mathbf{\mathsf{H}}$ of dimensions $(N\times N)$
which is invariant under the combined action of parity $\mathbf{\mathsf{P}}$
and time reversal $\mathbf{\mathsf{T}}$, Eq. (\ref{eq: PT invariance}), here is an algorithm qualitatively determines its qualitative spectrum:
\begin{enumerate}
\item  Calculate the characteristic polynomial $p_{\mathbf{\mathsf{H}}}(\lambda)$
of the matrix $\mathbf{\mathsf{H}}$;
\item  Determine the first $(2N-2)$ Newton sums $s_{n}$ associated
with the polynomial $p_{\mathbf{\mathsf{H}}}(\lambda)$;
\item  Write down the Hankel matrix $\mathfrak{H}_{\mathbf{\mathsf{H}}}$
 (\ref{eq: hankel of p}), defined in terms of the sums $s_{n}$;
\item  Obtain the number of constancies $\pi$ and alterations
$\nu$ in the sequence of signs (\ref{eq:signsequence}) giving the
inertia of $\mathfrak{H}_{\mathbf{\mathsf{H}}}$ as $\texttt{In}\,\mathfrak{H}_{\mathbf{\mathsf{H}}}=\{\nu,0,\pi\}$;
\item Then, the inertia of the $\mathbf{\mathsf{PT}}$-invariant matrix $\mathbf{\mathsf{H}}$
follows from the inertia $\texttt{In}\,\mathfrak{H}_{\mathbf{\mathsf{H}}}$
using (\ref{eq: inertia relations}) with $N\equiv\pi+\nu$, 
\begin{equation}
\texttt{In}_{\Re}\,\mathbf{\mathsf{H}}=\{\nu,N-2\nu,\nu\}\,.
\label{eq:spectral structure}
\end{equation}
\end{enumerate} 
Consequently, $\mathbf{\mathsf{PT}}$-symmetry is broken if $\nu>0$,
and $\mathbf{\mathsf{H}}$ will have $\nu$ pairs of complex conjugate
eigenvalues while the remaining $(N-2\nu)$ ones are real. Thu, the main 
result of this paper has been established.

\subsection{Example: The discretized $\mathbf{\mathsf{PT}}$-symmetric square well\label{sec:Example}}

\begin{figure}
\begin{center}
\psset{xunit=1mm,yunit=1mm,runit=1mm}
\psset{linewidth=0.3,dotsep=1,hatchwidth=0.3,hatchsep=1.5,shadowsize=1}
\psset{dotsize=0.7 2.5,dotscale=1 1,fillcolor=black}
\psset{arrowsize=1 2,arrowlength=1,arrowinset=0.25,tbarsize=0.7 5,bracketlength=0.15,rbracketlength=0.15}
\begin{pspicture}(0,0)(88,50)
\psline[arrowsize=2.25 2,arrowlength=1.1,arrowinset=0]{->}(8.37,29.77)(88,30)
\psline[arrowsize=2.25 2,arrowinset=0]{->}(8.37,9.88)(88,10)
\psline[arrowsize=2.25 2,arrowlength=1.1,arrowinset=0]{->}(45,15)(45,50)
\rput{0}(45,30){\psellipse[linewidth=0.2,fillstyle=vlines,hatchwidth=0.1,hatchsep=2.82,hatchangle=0](0,0)(10,10)}
\psline[linewidth=0.1,linestyle=dashed,dash=0.5 0.5](10,20)(80,40)
\psline[linewidth=0.1,linestyle=dashed,dash=0.5 0.5](10.45,36.42)(82,23)
\psline[arrowsize=2.25 2,arrowlength=1.1,arrowinset=0]{->}(82.21,44.19)(17,19)
\pscustom[linewidth=0.2,fillstyle=vlines,hatchwidth=0,hatchsep=2.2,hatchangle=-70]{\psbezier(82.56,40.23)(68.12,36.24)(58.74,33.28)(56.5,32)
\psbezier(56.5,32)(54.26,30.72)(54.62,29.46)(57.5,28.5)
\psbezier(57.5,28.5)(60.38,27.54)(69.23,25.65)(82.09,23.26)
}
\pscustom[linewidth=0.2,fillstyle=vlines,hatchwidth=0.1,hatchsep=2.2,hatchangle=-70]{\psbezier(10.58,20.58)(24.46,24.73)(31.26,27.04)(33.26,28.28)
\psbezier(33.26,28.28)(35.25,29.53)(35.43,30.36)(33.86,31.04)
\psbezier(33.86,31.04)(32.3,31.73)(28.52,32.6)(21.25,33.92)
\psbezier(21.25,33.92)(13.98,35.24)(10.87,35.83)(10.87,35.87)
}
\rput(81.98,17.79){$\xi$}
\psline[linewidth=0.1,linestyle=dashed,dash=0.5 0.5](55.12,29.94)(55.12,10)
\psline[linewidth=0.1,linestyle=dashed,dash=0.5 0.5](35.01,29.93)(35.01,9.99)
\rput(54.77,6.34){$\sqrt{2}$}
\rput(52.21,6.45){}
\rput(35.06,6.05){$-\sqrt{2}$}
\rput(45.12,12.44){$\mathbb{R} \times 3$}
\rput(21.28,12.44){$\mathbb{R} \times 2 \cup \mathbb{C} \times 2$}
\rput[l](60,12.56){$\mathbb{R} \times 1 \cup \mathbb{C} \times 2$}
\rput[r](41.86,46.16){$\Re E$}
\rput[l](22.67,18.95){$\Im E$}
\rput(82.21,5.47){$\xi$}
\rput[l](9.65,6.05){$(a)$}
\rput[l](9.77,42.67){$(b)$}
\end{pspicture}
\caption{Comparison of (a) the qualitative spectrum (\ref{eq: PT squ well inertia}) obtained algorithmically with (b) the exact eigenvalues of the Hamiltonian (\ref{eq: PT sym Hamiltonian}) describing the discretized PT-symmetric well}
\label{fig: exactvsalgevs}
\end{center}
\end{figure}
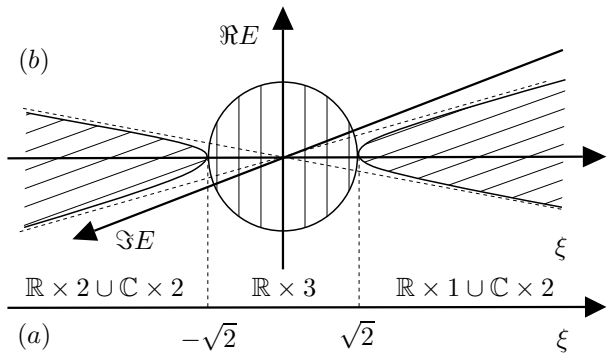

Let us work through the algorithm to detect the qualitative spectrum of the $\mathbf{\mathsf{PT}}$-symmetric
discretized square-well potential described by the matrix $\mathbf{\mathsf{H}}$ in (\ref{eq: PT sym Hamiltonian})---this time \emph{without} solving for its eigenvalues. The derivative of its characteristic polynomial (\ref{eq: charpoly Discrete SW}) reads
\begin{equation}
\frac{dp_{\mathbf{\mathsf{H}}}(\lambda)}{d\lambda}=3\lambda^{2}-(2-\xi^{2})\,.
\label{eq: cubic char poly}
\end{equation}
Compare the expansion 
\begin{equation}
\frac{p_{\mathbf{\mathsf{H}}}^{\prime}(\lambda)}{p_{\mathbf{\mathsf{H}}}(\lambda)}=3\lambda^{-1}+2(2-\xi^{2})\lambda^{-3}+2(2-\xi^{2})^{2}\lambda^{-5}+\mathcal{O}(\lambda^{-7})\,
\label{eq: newton expansion}
\end{equation}
with (\ref{eq: generate newton sums}), to read off the first five Newton sums. The Hankel matrix associated with $\mathbf{\mathsf{H}}$ is given by
\begin{equation}
\mathfrak{H}_{\mathbf{\mathsf{H}}}=2\left(\begin{array}{ccc}
3/2 & 0 & (2-\xi^{2})\\
0 & (2-\xi^{2}) & 0\\
(2-\xi^{2}) & 0 & (2-\xi^{2})^{2}\end{array}\right)\,,
\label{eq: explicit hankel}
\end{equation}
and its leading principal minors have determinants
\begin{equation}
d_{1}=3,\quad d_{2}=6(2-\xi^{2}),\quad d_{3}=20(2-\xi^{2})^{3}\,.
\label{eq: principal minors}
\end{equation}
Depending on the value of the parameter $\xi$, two different
sequences of signs arise: for $\xi^{2}<2$, one has all $d_{n}$
positive, resulting in \emph{three} constancies and \emph{no} alteration:
\begin{equation}
++++\quad\Rightarrow\quad\texttt{In}\,\mathfrak{H}_{\mathbf{\mathsf{H}}}=\{0,0,3\}\:,\label{eq: explicit signs a}
\end{equation}
while $d_{2}$ and $d_{3}$ turn negative for $\xi^{2}>2$,
implying that 
\begin{equation}
++--\quad\Rightarrow\quad\texttt{In}\,\mathfrak{H}_{\mathbf{\mathsf{H}}}=\{1,0,2\}\,.\label{eq:explicit sign b}
\end{equation}
Using the relation (\ref{eq:spectral structure}), the inertia of $\mathbf{\mathsf{H}}$ with respect to the
real axis is finally given by
\begin{equation}
\texttt{In}{}_{\Re}\mathbf{\mathsf{H}}=\left\{ \begin{array}{rcc}
\{0,3,0\} & \mbox{ if } & |\xi|<\sqrt{2}\, ,\\
\{1,1,1\} & \mbox{ if } & |\xi|>\sqrt{2}\, .\end{array}\right.
\label{eq: PT squ well inertia}
\end{equation}
Thus, the spectrum of $\mathbf{\mathsf{H}}$ is \emph{real} for $\xi^{2}<2$, while a pair of \emph{complex} eigenvalues exists whenever $\xi^{2}>0$. This agrees with the exact result (\ref{eq: Discret SW eigenvalues}) as depicted in Fig. \ref{fig: exactvsalgevs}.
For $\xi=\pm2$, the method cannot be applied since the
matrix $\mathfrak{H}_{\mathbf{\mathsf{H}}}$ in (\ref{eq: explicit hankel})
develops leading principal minors with vanishing determinant. This
is consistent with the fact that for these values of $\xi$ the properties of the matrix $\mathbf{\mathsf{H}}$ undergo qualitative changes such as the `disappearance' of an eigenstate. However, this does not put the current approach in jeopardy since these exceptional points can be identified \emph{beforehand} by running the algorithm presented in \cite{weigert06} which checks whether a given $\mathbf{\mathsf{PT}}$-invariant matrix is diagonalizable. In the present example, the points $\xi=\pm2$ would be flagged as shown explicitly in \cite{weigert2005}.

In fact, modifications of the current approach have been developed by Gundelfinger and Frobenius (cf. \cite{lancaster+85}) which are able to cope with the presence of at most \emph{three} consecutive \emph{vanishing} determinants $d_{n}$. The general relation between vanishing principal sub-determinants $d_{n}$ of the Hankel
matrix $\mathfrak{H}_{\mathbf{\mathsf{H}}}$ and the zeros of the
polynomial $p_{\mathbf{\mathsf{H}}}(\lambda)$ is not
obvious. In view of the example discussed above, it seems reasonable
to conjecture that there is a close link between the non-diagonalizability
of the matrix $\mathbf{\mathsf{H}}$ and the existence of vanishing
leading submatrices $d_{n}$ of $\mathfrak{H}_{\mathbf{\mathsf{H}}}$.

\section{Discussion and Outlook}

An algorithm has been presented which is capable to determine whether the eigenvalues of a $\mathbf{\mathsf{PT}}$-invariant
matrix $\mathbf{\mathsf{H}}$ (or possibly a family of such matrices depending smoothly on parameters) are complex or not. It complements an earlier algorithm \cite{weigert06} which detects whether a $\mathbf{\mathsf{PT}}$-invariant matrix does have a complete set of eigenstates. Thus, the fundamental questions \textbf{Q1} and \textbf{Q2} about $\mathbf{\mathsf{PT}}$-invariant systems spelled out in the introduction can be answered in a systematic
way if the system is described by a matrix of \emph{finite} but arbitrarily large dimension.

Although desirable, it is not obvious how to generalize the algorithm presented here to operators such as $\mathbf{\mathsf{H}}=p^{2}+ix^{3}$, acting in an \emph{infinite-dimensional} space. This observation
also applies to the algorithm for diagonalizability. From a numerical point of view, the algorithm is not particularly efficient since a total of $2N$ determinants of order up to $N$ need to be calculated. However, the method proposed here is \emph{exact}, contrary to any numerical implementation which would directly calculate the (approximate) eigenvalues of the matrix $\mathbf{\mathsf{H}}$. More efficient algorithms to determine the spectrum of a $\mathbf{\mathsf{PT}}$-invariant
matrix are likely to exist---Sturmian sequences based on the Euclidean algorithm for polynomials \cite{weigert(inprep)} being the most promising candidates. 

\bibliographystyle{unsrt}

\end{document}